\documentclass[submission, Phys]{SciPost}
\usepackage[american]{babel}
\usepackage{bm}
\usepackage{graphicx}
\usepackage{amsmath}
\usepackage{amsfonts}
\usepackage{amssymb}
\usepackage{amsthm}
\usepackage{amstext}
\usepackage{amsbsy}
\usepackage{amsopn}
\usepackage{amscd}
\usepackage{amsxtra}
\usepackage{sidecap}

\newcommand{\Mc}[1]{\mathcal{#1}}

\newcommand{\setN}{\mathbb{N}}
\newcommand{\setR}{\mathbb{R}}

\newcommand{\setC}{\mathbb{C}}
\newcommand{\Id}{\mathbb{I}}

\newcommand{\bra}[1]{\langle #1 |}
\newcommand{\ket}[1]{| #1 \rangle }
\newcommand{\braket}[2]{\langle #1| #2 \rangle }
\newcommand{\ii}{\textsl{i}}
\newcommand{\A}{\hat a}
\newcommand{\ad}{\hat a^\dagger}

\begin{document}


\begin{center}{\Large \textbf{
Loop quantum gravity with optimal control path integral, and application to black hole tunneling}}
\end{center}

\begin{center}
Q. Ansel\textsuperscript{*}
\end{center}

\begin{center}
Institut UTINAM - UMR 6213, CNRS, Universit\'{e} Bourgogne Franche-Comt\'{e}, Observatoire des Sciences de l'Univers THETA, 41 bis avenue de l'Observatoire, F-25010 Besan\c{c}on, France
\\
* quentin.ansel@univ-fcomte.fr
\end{center}

\begin{center}
\today
\end{center}

\begin{abstract}
{\bf
This paper presents a novel path integral formalism for Einstein's theory of gravitation from the viewpoint of optimal control theory. Despite its close relation to the well-known variational principles of physicists, optimal control turns out to be more general. Within this context, a Lagrangian different from the Einstein-Hilbert Lagrangian is defined. Einstein field equations are recovered exactly with variations of the new action functional. The quantum theory is obtained using Ashtekar variables and the loop scalar product. By means of example, the tunneling process of a black hole into another black hole or into a white hole is investigated with a toy model. 
}
\end{abstract}

\vspace{10pt}
\noindent\rule{\textwidth}{1pt}
\tableofcontents\thispagestyle{fancy}
\noindent\rule{\textwidth}{1pt}
\vspace{10pt}

\section{Introduction}

The Lagrangian and Hamiltonian formulation of general relativity is a long-standing problem that has been solved mostly in the sixties by Dirac, Arnowit, Deser, Misner \cite{dirac_theory_1958,arnowitt_dynamics_1962,misner_gravitation_1973}, and in the eighties by Ashtekar \cite{ashtekar_new_1986} for the connection formulation. Most of the scientific research on this branch of physics has followed the path sketched by these papers (see \cite{rovelli_notes_2001} for a historical review). The problem could have been considered closed, but remaining technical difficulties encountered in quantum gravity (such as the difficulty to properly determine the Hamiltonian operator in Loop Quantum Gravity (LQG) \cite{rovelli_quantum_2007,wieland_complex_2012,yang_new_2015}, or the difficulties encountered with spinfoam \cite{rovelli_covariant_2014,perez_spin-foam_2013,vojinovic_cosine_2014,dona_infrared_2018,christodoulou_divergences_2013,dona_numerical_2018,dona_numerical_2020,dona_searching_2020}) leads us to think that maybe, this is not the end of the story, and a novel starting point is required. For example, we can modify the variational principle, as suggested by  C. Cremaschini, M. Tessarotto in Refs.~\cite{cremaschini_synchronous_2015,cremaschini_manifest_2016,cremaschini_hamiltonian_2017,cremaschini_quantum-wave_2017}.

Independently of this context, a Lagrangian and Hamiltonian formalism for optimal control problems have been developed during the second half of the 20th century, by Bellman, Pontryagin, and their collaborators \cite{pesch_maximum_2009,kirk_optimal_2004,trelat_controoptimal:_2005}. Since, Optimal Control Theory (OCT) has been one of the most successful theories of mathematics, with applications in engineering, aerospace \cite{bonnard_optimal_2012}, robotics, finance, quantum technologies \cite{glaser_training_2015,boscain_introduction_2020},... Despite its close relation to the well-known variational principles used in classical physics, optimal control has a few small differences that allows us to tackle more more general situations \cite{contreras_dynamic_2017}. In particular it can handle dynamical problems without natural canonical adjoint state of the generalized coordinates. More recently, several papers have outlined the precise relationship between OCT and classical/quantum physics \cite{kime_two_1989,contreras_dynamic_2017,koppe_derivation_2017,ohsumi_interpretation_2019,guerra_quantization_1983}. As an example, one can explain how quantum mechanics can be understood through stochastic optimization on space-times \cite{lindgren_quantum_2019,papiez_stochastic_1982}.

Following the idea that a novel approach is required to solve quantum gravity dynamics, the issue is explored using the optimal control formalism. The first step consists of rewriting general relativity dynamics in a suitable form. This step can be performed using any variables in 3+1 formalism. Then, a Lagrangian and a Hamiltonian can be defined over an extended configuration space. It turns out that the corresponding path integral propagator takes a very simple form. However, optimal control does not give us a natural way to construct quantum gravitational states. We thus employ Ashtekar variables and the LQG scalar product for that purpose. By means of an example, we consider a toy model of black hole tunneling. This hypothetical quantum effect \cite{christodoulou_planck_2016,dambrosio_end_2020} has been studied recently in the context of LQG in order to shed light on Planck stars \cite{rovelli_planck_2014,barrau_planck_2014}.

This article is structured as follows, in Sec.~\ref{sec:sota_OCT} the relation between optimal control and physical systems is reviewed. In Sec.~\ref{sec:Path integral quantization} the path integral quantization procedure is defined and analyzed with the harmonic oscillator. In Sec.~\ref{sec:GR} and Sec.~\ref{sec:path_integral_GR} we apply the theory to general relativity expressed in Ashtekar variables and the path integral is derived. The application to black hole tunneling is considered in Sec.~\ref{sec:BH_tunneling}. Finally, a conclusion is made in Sec.~\ref{sec:conclusion}.

\section{Lagrangian and Hamiltonian in OCT}
\label{sec:sota_OCT}

This first section is devoted to a concise presentation of Lagrangian and Hamiltonian formalism in optimal control theory \cite{kirk_optimal_2004,bonnard_optimal_2012,trelat_controoptimal:_2005}. To illustrate the subject, an application to the harmonic oscillator is presented. For further technical details, concerning the relations between standard physics and optimal control, we refer to \cite{contreras_dynamic_2017}.

We consider a physical system described by the real variable $x\in \Mc C$ (a classical state that belongs to the configuration space) whose dynamics are governed by the first order differential equation: $\dot x = f(x,u(t),t)$. $u(t)$ is a time-dependent control field, an input of the system that must be determined. A standard optimal control problem is to determine $u$ in order to transform $x$ from an initial state $x_0$ to a target state $x_t$ while minimizing some constraints (the time of the transformation, the energy consumption,...). To solve this problem, a Lagrangian and a Hamiltonian are constructed, by analogy to standard classical mechanics. Interestingly, it allows us to describe situations ill-defined in classical mechanics. For a given control problem, we define the cost function:
\begin{equation}
C=\int_0^t dt' f_0(x,u,t')
\end{equation}
Extremums of t$C$ are found using variational calculus. However, to take into account the constraints imposed by the system dynamics, we introduce a dynamical constraint with the use of a Lagrange multiplier. We define:
\begin{equation}
S=\int_0^t dt' (f_0(x,u,t') + p_x(\dot x - f(x,u,t')) + p_u).
\end{equation}
Which can be assimilated to the physical action, and $\Mc L_{OC} = f_0(x,u,t') + p_x(\dot x - f(x,u,t')) + p_u$ is the system Lagrangian in the extended configuration space $\Mc C'$ defined by the vector $(x,u,p_x,p_u)$. Here, $p_u$ is a Lagrange multiplier for the control field. For a constrained control field, $p_u $ must be multiplied by a function of $u$, $\dot u$,  or any parameter that specifies the constraints on the control field. For simplicity, we consider here an unconstrained control. Extremals of this action are described by Euler-Lagrange's equation, $\partial_X \Mc L_{OC} - d_t (\partial_{ \dot X} \Mc L_{OC})=0$, with $X \in \{x,u,p_x,p_u\}$. For the system above, one obtains: 
\begin{align}
\dot x &= f(x,u,t) \\
\dot p_x &= \partial_x (f_0 - p_x f) \\
0 & = \partial_u (f_0 - p_x f)) \\
p_u & = 0
\end{align}
We can also define the Pontryagin Hamiltonian:
\begin{equation}
H_{OC} = p_x \dot x + p_u \dot u - \Mc L_{OC}
\end{equation}
Hamilton's equations $\partial_X H_{OC}= - \dot P$ and $\partial_P H_{OC}=  \dot X$, $X\in\{x,u\}$ and $P\in\{p_x,p_u\}$ gives us again the equation of motions. From Hamilton equation, one also obtains:
\begin{equation}
\frac{\partial H_{OC}}{\partial u} = \frac{\partial}{\partial u} (p_x f - f_0 ) =0.
\end{equation}
This corresponds to Pontryagin Maximum Principle for an unconstrained control field. Notice that contrary to classical mechanics, the Hamiltonian is a function on $\Mc C'=(x,u,p_x,p_u)$, and the Lagrangian is a function of $(x,\dot x, u, \dot u, p_x, \dot p_x, p_u, \dot p_u)$. 

By means of an example, we consider the case of a harmonic oscillator, described by the following equations of motion:
\begin{equation}
\dot x = \frac{p}{m} ~; ~\dot p = - k x.
\label{eq:dyn_H.O.}
\end{equation}
where $x$ gives the position of the oscillator of mass $m$, and $p$ is its momentum. $k$ is the "spring constant". The configuration space is the phase space $\Mc C=(x,p)$. To define the optimal control Lagrangian and Hamiltonians, we define the extended configuration space $\Mc C'=(x,p,\chi,\pi)$, where $\chi$ and $\pi$ are assimilated to the respective adjoint states of $x$ and $p$. With this system, there is no control field and no additional constraints on the system dynamics. Then, the Lagrangian simply read:
\begin{equation}
\Mc L_{OC} = \chi \left(\dot x - \frac{p}{m}\right) + \pi \left( \dot p + kx \right),
\end{equation}
and the Hamiltonian is:
\begin{equation}
H_{OC} = \chi \frac{p}{m} - k \pi  x .
\end{equation}
using Hamilton's equations, we can deduce dynamical equations for the adjoint states:
\begin{equation}
\dot \pi = \frac{\chi}{m} ~; ~\dot \chi = - k \pi.
\end{equation}
They are the same as equations \eqref{eq:dyn_H.O.}. If we impose as initial conditions $\pi(0) = x(0)/2$ and $\chi(0)= p(0)/2$, the Hamiltonian becomes:
\begin{equation}
H_{OC} = \frac{1}{2}\left( \frac{p^2}{m} + k x^2 \right).
\end{equation}
This is the system's Hamiltonian in classical mechanics. With the optimal control approach, we have a more general description of the system's time evolution in terms of variations of a scalar function. Because the extended configuration space $\Mc C'$ is bigger than the phase space, there are many possible trajectories. However, they are all identical when they are projected on $\Mc C$. Only trajectories given by the equations \eqref{eq:dyn_H.O.} are physically possible. We shall now discuss the path integral quantization.

\section{Path integral quantization}
\label{sec:Path integral quantization}

Given a classical system described by the optimal control problem described in Sec.~\ref{sec:sota_OCT}, we can switch to a quantum theory by introducing a path integral. Remark that this is unrelated to stochastic optimal control where path integrals are also used.

The path integral of the optimal control problem is formally given by \cite{contreras_dynamic_2017}:
\begin{equation}
\hat W = \int \Mc D x \Mc D p_x e^{\ii \int_0 ^t \Mc L_{OC} dt'}.
\label{eq:formal_path_integral}
\end{equation}
Here, $x$ refers to the classical state, and $p_x$ to its adjoint state in $\Mc C'$. Since the control field is treated as a position coordinate, we do not include it in the equations. It is straightforward to recover the case with an explicit control field. If the control is classical, we do not integrate over all control possibilities.

The path integral \eqref{eq:formal_path_integral} can be tackled with standard tools of physics or optimal control \cite{feynman_quantum_2010,yang_path_2014}. To ensure a well-defined path integral, time is discretized. Then, the formal definition becomes:
\begin{equation}
\hat W = \Mc N \int \prod _t d x^{(t)} d p_x^{(t)} e^{\ii \sum_t \Delta t (f_0^{(t)} + p_x^{(t)} (\dot x ^{(t)} - f^{(t)})) },
\label{eq:lattice_path_integral}
\end{equation}
where $\Mc N= (\sqrt{2 \pi})^{-d}$ is a normalization coefficient, $\dot x^{(t)}, f_0^{(t)}$, and $f^{(t)}$ are functions of the system state at other time steps: $x^{(t-1)},p^{(t-1)},x^{(t-2)},p^{(t-2)}$, etc. For example, if one discretizes derivatives with an Euler method, $\dot x^{(t)} = \frac{1}{\Delta t}(x^{(t)}-x^{(t-1)})$. 

We notice in Eq.\eqref{eq:lattice_path_integral} that adjoint states are used only once, at each time step. Then the integration over $p_x^{(t)}$ is simple:
\begin{equation}
\hat W = \int \prod _t d x^{(t)}  \delta_0 \left( \Delta t (\dot x ^{(t)} - f^{(t)}) \right) e^{\ii \sum_t \Delta t f_0^{(t)}}.
\label{eq:path_integral_delta}
\end{equation}
The adjoint state concentrate the possible paths along classical trajectories. Since there is only one Dirac distribution per $x^{(t)}$, this last integral is well defined. Additionally it is a simple propagator of the system dynamics, as we can expect in the original theory of Faynman.

A simple application can be made using the harmonic oscillator. The quantum state is a function $\psi (x,p)$. If at time $t=0$ the system is in the state $\psi (x^{(0)},p^{(0)})$, at time $\Delta t$ we have from Eq.~\eqref{eq:path_integral_delta}:
\begin{equation}
\begin{split}
&\int dx^{(0)} dp^{(0)} \delta_0\left( x^{(1)}-x^{(0)} - \Delta t \frac{p^{(0)}}{m}\right) \delta_0 \left(p^{(1)}-p^{(0)} + \Delta t k x^{(0)}\right) \ket{\psi (x^{(0)},p^{(0)})} \\
&
= \left(1+\frac{k \Delta t^2}{m} \right) \ket{\psi \left( \frac{m x^{(1)- \Delta t p^{(1)} }}{m + k \Delta t^2}, m\frac{p^{(1)}+ k \Delta t x^{(1)}}{m+k\Delta t ^2}\right)}
\end{split}
\end{equation}
To be consistent with the first order approximation, we have $\Delta t^2 \simeq 0$. Then, the last equation becomes:
\begin{equation}
\ket{\psi\left(x^{(1)}- \frac{\Delta t}{m} p^{(1)}, p^{(1)} + k \Delta t x^{(1)} \right)}
\end{equation}
It is therefore obvious that the propagator has the following property:
\begin{equation}
\hat W(t) \ket{\psi (x,p)} = \ket{\psi(\hat W^{-1}(t)x(t),\hat W^{-1}(t)p(t))},
\label{eq:evolution_operator_from_path_integral}
\end{equation}
where $\hat W^{-1}(t) X$ denotes the propagation in inverse time of $X$, using classical equations of movement.

Then, solving the time evolution of the wave function is a trivial problem once the flow of the classical state is known. This can be solved analytically for the simplest systems, but one shall use numerical integrations for non-linear systems.

At this point, the main question concerns the consistency of this theory with the standard theory of quantum mechanics. We propose to explore this issue with the example of the harmonic oscillator \cite{gardiner_quantum_2004}. The standard quantum Hamiltonian operator is (in $\hbar$ units): $\hat H = \omega \ad \A$, with eigen-values $\omega n$ and eigen-states $\ket{n}$. To make the relation with the path integral quantization, we shall introduce the coherent state $\ket{\alpha} = e^{-|\alpha|^2/2}\sum_{n=0}^\infty \frac{\alpha^n}{\sqrt{n!}} \ket{n}$. A straightforward calculation of the evolution operator $\hat U$ applied to an arbitrary coherent state $\ket{\alpha_0}$ gives:
\begin{equation}
e^{-\ii t \hat H}\ket{\alpha_0} = \ket{\alpha_0 e^{-\ii \omega t}}.
\end{equation}
Then,
\begin{equation}
\ket{\alpha(t)} = \hat U(t) \ket{\alpha_0}= \ket{\alpha_0 e^{-\ii \omega t}}.
\label{eq:evolution_operator_coherent_state}
\end{equation}
It must be noted that $\alpha$ is the classical counterpart of the annihilation operator $\A$. Therefore, it is related to $x$ and $p$ by the relation:
\begin{equation}
\alpha = \frac{\ii}{\sqrt{2 \omega m}} p + \sqrt{\frac{k}{2 \omega}} x,
\label{eq:def_alpha}
\end{equation}
and it follows the classical equations of motion \cite{gardiner_quantum_2004}:
\begin{equation}
\frac{d\alpha}{dt} = -\ii \omega \alpha.
\label{eq:classical_EOM_alpha}
\end{equation}
The solution of this equation is simply $\alpha(t) = e^{-\ii \omega t} \alpha_0$. From these definitions, it is clear that $\ket{\alpha}$ corresponds to a state $\ket{\psi(x,p)}$ with the mapping $(x,p)\rightarrow \alpha$. Moreover, we have $\alpha_0 = e^{\ii \omega t} \alpha(t)$, which corresponds to $\hat W^{-1}(t) \alpha(t)$ in Eq.~ \eqref{eq:evolution_operator_from_path_integral}. The only difference with \eqref{eq:evolution_operator_coherent_state} is that the initial state is used to parameterize the system time evolution. It turns out that the path integral propagator of the optimal control formulation of the harmonic oscillator is equivalent to the standard evolution operator.

Also, the path integral must encodes the scalar product between two coherent states. From now, nothing tells us what the scalar product is. This issue can be solved with the introduction of a boundary/terminal term inside the cost function, which does not influence the dynamics. In OCT, the terminal cost is usually the distance between the final state and the target state (see \cite{werschnik_quantum_2007} for an application of OCT to controlled quantum systems). It is a measure of distance in the configuration space.  The boundary terms for the harmonic oscillator can be inferred as follows: the scalar product between two coherent states is $\braket{\alpha}{\beta} = \exp(-|\alpha|^2/2 -|\beta|^2/2 + \alpha^* \beta)$. We notice that when the probability is computed we get $|\braket{\alpha}{\beta}|^2 = \exp(|\alpha|^2 -|\beta|^2 + 2 \Re( \alpha^* \beta)) = \exp(-|\alpha - \beta|^2)$. This is the exponential of the square distance between $\alpha$ and $\beta$. Therefore, the terminal cost is given by:
\begin{equation}
S_B = -\ii (-|\alpha|^2/2 -|\beta|^2/2 + \alpha^* \beta)
\end{equation}
with $\alpha$, $\beta$, defined by Eq.~\eqref{eq:def_alpha}.

\section{Optimal control formulation of general relativity}
\label{sec:GR}

In the previous section, a simple path integral was derived using optimal control theory. The idea is to proceed similarly for gravity, and to develop an alternative to spinfoams. 

Contrary to the standard approach, the idea is not to assume a Lagrangian that must provide Einstein's equation using some variational rules, but a Lagrangian is constructed from Einstein's field equations with an optimal control approach.
As another guideline the theoretical proposal  must be computationally efficient (at least, by using state of the art numerical methods in quantum mechanics and relativity). 

\subsection{Optimal control action for general relativity}

Einstein field equations (in the vacuum) are generally written as \cite{misner_gravitation_1973}:
\begin{equation}
R_{\mu \nu} + g_{\mu \nu }\left( \Lambda - \frac{1}{2} R\right) = 0.
\label{eq:einstein_equation}
\end{equation}
%
%
$R_{\mu \nu}$ is the Ricci tensor, calculated using the metric tensor $g_{\mu \nu}$. $\Lambda$ is the cosmological constant, and $R = R_{\mu\nu}g^{\mu \nu}$.
Written in this covariant form, Einstein's field equations are not easy to handle: there is no explicit notion of evolution, there are redundancies in the variables, and some gauge degrees of freedom must be chosen in order to make explicit calculations. 
In order to make an easier link with state of the art canonical quantum gravity (i.e. the standard loop approach), we introduce Ashtekar variables $E_i^a$ and $A^i_a$ \cite{ashtekar_new_1986,barbero_g_real_1995,pons_gauge_2000,rovelli_quantum_2007,rovelli_covariant_2014}. However, the following theory is sufficiently flexible to use any set of variables, such as ADM variables, or any other variables introduced in numerical relativity \cite{baumgarte_numerical_2010,bardeen_tetrad_2011,hamilton_covariant_2017}. 

To derive Ashtekar variables, it is necessary to express the metric with a tetrad field:
\begin{equation}
g_{\mu \nu} = e_\mu ^I e_\nu^J \eta _{IJ},
\end{equation}
with $ \eta _{IJ}$ the Minkowski metric, and $e_\mu^I$ the tetrad. In the following, we use the signature $(-,+,+,+)$. The next step is to decompose the tetrad as follows \cite{pons_gauge_2000}:
\begin{equation}
e_{\mu}^I = \left( \begin{array}{cc}
\mathsf{N} & 0 \\ 
e^i_a \mathsf{N}^a &  e_a^i
\end{array} \right),
\end{equation}
with $\mathsf{N}$, $\mathsf{N}^a$ the lapse function and the shift vectors, which are gauge degrees of freedom. $e^a_i$ is the triad. Finally, we define the "gravitational electric field":
\begin{equation}
E_i^a = \det (e^i_a) e_i^a,
\end{equation}
and the extrinsic curvature:
\begin{equation}
k^i_a = \frac{e^{bi} }{2\mathsf{N}} \left( \partial_t (e_a^j e_b^k)\delta_{jk} + D_{(a}\mathsf{N}_{b)} \right).
\end{equation}
The operator $D_a$ is the covariant derivative of the three metric. To finalize the definition of Ashtekar variables, we introduce the gravitational potential:
\begin{equation}
A_a^i = \omega_a ^i + \beta k_a^i,
\end{equation}
where $\omega_a^i = \omega_a^{jk}\epsilon^i_{~jk}$ is the triad spin connection and $\beta$ is a real parameter (it is also possible to define the theory with a complex parameter).  Dynamics of $E_i^a$ and $A^i_a$ are given by first order differential equations. For conciseness, we give here the differential equations for $\Lambda = 0$, and in the gauge $\mathsf{N}=1$, $\mathsf{N}_a = 0$ (further details on the differential equations, for $\beta = \ii$ can be found in~\cite{gambini_loops_1996,shinkai_hyperbolic_2000}).
\begin{equation}
\dot E^a_i =  \frac{1}{\beta \sqrt{\det(E^i_a)}} \left(A^j_b - \omega^j_b \right)(E^b_jE_i^a - E^b_i E^a_j)
\label{eq:dot_E}
\end{equation}
\begin{equation}
\dot A^i_a =   \frac{1}{2 \beta \sqrt{\det(E^a_i)}}  E^b_j \epsilon^{ijk }F_{abk}
\label{eq:dot_A}
\end{equation}
%
%
We have introduced $F_{abk}=F_{ab}^i \delta_{ik}$ the curvature two forms of $A_a^i$. In the following, the explicit formula of each time derivative is of little interest. We only require their existence, such that the Cauchy problem for general relativity is well posed. In addition to these dynamical equations, the gravitational field must verify some constraints:
\begin{align}
\label{eq:C_0}
C_0 & = \Mc E_{\partial \Sigma} + E^a_i E^b_j \epsilon^{ijk }F_{abk} 
 - \frac{2(\beta ^2 +1)}{\beta^2} E^a_{[i}E^b_{j]}(A^i_a - \omega^i_a)(A^j_b - \omega^j_b) =0, \\
\label{eq:C_a}
 C_i & = D_a E_i^a = 0, \\
\label{eq:C_i}
 C_a & = -F_{ab}^i E_i^b = 0.
\end{align}
We introduced $\Mc E_{\partial \Sigma}$, which is a possible surface contribution in the system energy \cite{regge_role_1974}.
Equations \eqref{eq:dot_E}, \eqref{eq:dot_A} \eqref{eq:C_i}, \eqref{eq:C_a}, and \eqref{eq:C_0} must be verified at each point of the space-time manifold. Hence, we have to take into account this fact in the construction of the action functional.

In order to obtain a well-defined variational principle which is compatible with optimal control theory, we propose the following construction: we choose a finite number of $N$ points in a 3D-hypersurface $\Sigma$ of the space-time manifold. They define a network in $\Sigma$, noted $X_N$. A sequence of networks is defined by increasing successively the number of points such that $ X_N < X_{N+1}$ (the first $N$ points of $X_{N+1}$ are identical to the points of $X_N$, and the point $N+1$ is a new point in the network). At each point of $X_N$, we have a couple $(E_i^a,A^i_a)$ that  evolve along a world-line; which is parametrized by the time variable $t$. For this set of points, the following action is defined:

\begin{equation}
\begin{split}
S_N = \sum_{n=1}^N \int_{t_0}^{t_f}dt~& P_a^i(x_n)(\dot E_i^a (x_n) - \Mc G_i^a (x_n) ) \\
&+ \Pi^a_i(x_n) (\dot A_a^i(x_n) - \Mc F_a^i (x_n) ) \\
& + \lambda^A (x_n) C_A (x_n).
\end{split}
\label{eq:action_GR_no_boundary_v0}
\end{equation}
where $\Mc G$ and $\Mc F$ are given respectively by the right side of \eqref{eq:dot_E}, and \eqref{eq:dot_A}, $P_a^i$, $\Pi_i^a$, and $\lambda^A$ are adjoint states, and the indice $A$ takes the values $(i,a,0)$.

When the number of points $x_n$ covers $\Sigma$ sufficiently well, such that the covariant derivative $D_a$ can be computed using the field variables of the neighborhood points (with finite differences, for example), we can define a continuous action functional on $\Sigma$. Let $\Mc P_\Sigma X_N$ be a partition of $\Sigma$ such that each cell is unequivocally associated with a single point of $X_N$. Let $\Mc V_N = \max_{\sigma \in \Mc P_\Sigma X_N}(\text{Vol} (\sigma))$, with $\text{Vol} $ the Lebesgue measure. When $\lim_{N \rightarrow \infty} \Mc V_N = 0$ and when the discretized versions of $\dot E_i^a,\dot A^i_a, \Mc G^a_i, \Mc F_a^i,C_A$ converge in measure toward the continuous versions, we define the action of the full space-time manifold:
\begin{equation}
S= \lim_{N \rightarrow \infty} S_N,
\end{equation}
and we have:
\begin{equation}
\begin{split}
S = \int_{t_0}^{t_f}dt \int_\Sigma d^3x ~& P_a^i(x)(\dot E_i^a (x) - \Mc G_i^a (x) ) \\
&+ \Pi^a_i(x) (\dot A_a^i(x) -  \Mc F_a^i (x) ) \\
& + \lambda^A (x) C_A (x).
\label{eq:action_GR_no_boundary}
\end{split}
\end{equation}
Remark that the adjoint states written in Eq.~\eqref{eq:action_GR_no_boundary} are in fact proportional to the ones in Eq.~\eqref{eq:action_GR_no_boundary_v0} by a factor $\text{Vol}(\sigma)^{-1}$ in order to use functional derivatives in Euler-Lagrange equations instead of partial derivatives. Moreover, the field variables are desensitized quantities, and therefore, an integration weight $\det (e^i_a)$ is implicit in Eq.~\eqref{eq:action_GR_no_boundary}. Notice that it is not excluded that $S= \pm\infty$ for some configurations, but $S=0$ for any solution of Einstein field equations.

From \eqref{eq:action_GR_no_boundary}, we can determine the equation of motions for the adjoint variables $P_a^i, \Pi _i ^a , \lambda^A$. We do not present the them here because tthey do not play a role in what follows. We left the study of the adjoint states dynamics, and the study of the Pontryagin Hamiltonian to another study. They could have an interesting geometrical interpretation and a possible impact on the canonical quantization. A few preliminary results in this direction are given in Appendix \ref{sec:hamiltonian operator}.

\subsection{The gravitational boundary term: the quantum scalar product}
To finish with the action functional, we must provide the boundary term that encodes the distance in the space of physical solutions. There is a liberty of choice of definition, but there are also many physical constraints, imposed by covariance and gauge invariance. For this study we use the well-developed scalar product between two coherent spin-network states \cite{rovelli_covariant_2014,bahr_gauge-invariant_2009,stottmeister_coherent_I_2016,stottmeister_coherent_II_2016,stottmeister_coherent_III_2016,bianchi_coherent_2010}.
This allows us to make an explicit link with loop quantum gravity.

The usual approach is to define a spin-network state that describes a classical discretized manifold, by using complexifier coherent states. To each node of the spin-network we associate a cell of the discretized manifold. A link of the network corresponds to the boundary between two cells. In many cases it is convenient to work with simplicial manifold. In this case, a node of the spin network is 4-valent (it has four links), and it is associated with a tetrahedron. Here, we prefer to work with 6-valent nodes in order to describe hexahedrons. This choice allows us to make a simpler link with the classical geometry, defined by a metric tensor. Then, the classical manifold is discretized using hexahedrons, and the dual graph is used to define the spin-network. To each link of the network, we associate two quantities: 

\begin{itemize}
\item The holonomy of the Ashtekar connection between the centers of each hexahedron noted $x_s$ and $x_t$ for respectively the source and the target points.
\begin{equation}
U_l (x_s,x_t) = \mathbb{P}\exp\left( \frac{\ii}{2} \int_{x_s}^{x^t}  A_i^a(x) \hat \sigma^i t_a(x) dx \right) \approx \exp \left( \frac{\ii}{2} A_i^a(x_s) \hat \sigma^i  t_a(x_s) L \right),
\label{eq:holonomy}
\end{equation}
with $\mathbb{P}$ the path-ordering operator, $\hat \sigma^i$ the Pauli matrices, $t_a$ the unit tangent covector to the link, and $L$  the length of the path.

\item The second quantity is the integral of the "electric" field over the surface dual to the link:
\begin{equation}
X^i = \int_S \tilde E^i_a dS^a \approx U_l(x_l,x_s)\left[ E^i_a (x_l) n^a S \right] U_l^{-1}(x_l,x_s) .
\label{eq:Xi_surface}
\end{equation}
The tilde above $E^i_a$ denotes the fact that it is defined by the parallel transport of the electric field to the starting point $x_s$ of the link ~\cite{bahr_gauge-invariant_2009}. In the discretized version of this quantity, this is achieved by an holonomy $U_l (x_l,x_s)$, where $x_l$ is the intersection point between the path associated with the link and the surface. $n^a$ is the normal vector of the surface with area $S$.
\end{itemize} 
From these two quantities, we define the following $SL(2,\setC)$ matrix~\cite{bahr_gauge-invariant_2009}:
\begin{equation}
g_l = \exp\left(\frac{1}{2} X^i \hat \sigma_i \right)  U_l.
\label{eq:def_matrix_gl}
\end{equation}
The matrices associated with the links are used to parameterize the coherent state. The state is defined in a two-step procedure. First we define a function called "heat kernel":
\begin{equation}
K ^\zeta_{g_l} (U_l) = \sum_{j_l \in \setN/2} e^{-\zeta j_l (j_l +1)/2} d_{j_l} \text{Tr}^{j_l}(g_l^{-1} U_l),
\label{eq:def_heat_kernal}
\end{equation}
with $d_{j_l} \equiv 2 j_l +1$, $\text{Tr}^{j_l}$ is the trace in the spin-$j_l$ representation, and $\zeta$ is a parameter defining the coherent state. Semi-classical properties are obtained when $\zeta \rightarrow 0$. The second step is to take the product over the links of the heat kernel function, and to make the state gauge invariant at the nodes by integrating over $SU(2)$ at each node:
\begin{equation}
\psi^\zeta_{[ g_l ]} (U_1,...,U_{N_L}) = \int_{SU(2)^{N_N}} \prod_{c=1}^{N_N}dh_c \prod _{l=1}^{N_L} K ^\zeta_{h_a g_l h_b^{-1}} (U_l)
\label{eq:def_coherent_spin_network}
\end{equation}
where we have assumed that the spin network has $N_N$ nodes and $N_L$ links. In the integral, an $SU(2)$ element $h_c$ is associated to the node $c$. For convenience, we note $h_a$ the $SU(2)$ element for the source node of the link, and $h_b$ the $SU(2)$ element for the target node. Notice the use of the notation $[g_l]$, to specify that the state is gauge-invariant.

When $\zeta \rightarrow 0$, the coherent state has a Gaussian-like distribution over the spins $j_l$. The center of the Gaussian tends to infinity when $\zeta \rightarrow 0$. Then, for a sufficiently small $\zeta$, we can make a large $j$-approximation of the state. We review here the main idea of the approximation, but additional details can be found in \cite{bianchi_coherent_2010}. The idea is to rewrite the $SL(2,\setC)$ matrices $g_l$ defined in Eq.~\eqref{eq:def_matrix_gl} using a Cartan decomposition $g_l = u_l.e^{r_l \sigma_z/2}.v_l^{-1}$, with $r_l\in \setR_+$ and $u_l, v_l \in SU(2)$. Then, the heat kernel defined in Eq.~\eqref{eq:def_heat_kernal} can be rewritten as:
\begin{equation}
\begin{split}
K ^\zeta_{h_a g_l h_b^{-1}} (U_l) &= \sum_{j_l \in \setN/2} e^{-\zeta j_l (j_l +1)/2} d_{j_l} \text{Tr}^{j_l}(h_a g_l^{-1} h_b^{-1} U_l) \\
&= \sum_{j_l \in \setN/2} e^{-\zeta j_l (j_l +1)/2} d_{j_l} \sum_{m n l} D^{j_l}_{m n}(h_a u_l) D^{j_l}_{n l}(e^{r \sigma_z/2}) D^{j_l}_{lm}(v_l^{-1} h_b^{-1} U_l)
\end{split}
\label{eq:derivation_heat_kernel_approximation}
\end{equation}
In the large spin limit, we have $D^j_{nl}(e^{r \sigma_z/2}) \approx e^{r j}\ket{j}\bra{j}$ (since $r \sigma_z/2$ is a real diagonal matrix, the Wigner matrix is characterized by an ensemble of exponential, which is dominated by the largest eigenvalue).
\begin{equation}
K ^\zeta_{h_a g_l h_b^{-1}} (U_l) \approx \sum_{j_l \in \setN/2} e^{-\zeta j_l (j_l +1)/2} d_{j_l} D^{j_l}_{j_l j_l}(v_l^{-1} h_b^{-1} U_lh_a u_l)
\end{equation}
By introducing this equation into Eq.~\eqref{eq:def_coherent_spin_network}, we deduce:
\begin{equation}
\psi^\zeta_{[g_l]}(U_1,...,U_{N_L}) \approx \sum_{j_l} \prod_{l=1}^{N_L} e^{-\zeta j_l (j_l +1)/2} d_{j_l} e^{r_l j_l}\Psi_{j_l, u_l, v_l} (U_l),
\label{eq:approx_coherent_state}
\end{equation}
where $\Psi_{j_l, u_l, v_l} (U_l)$ is the intrinsic coherent state \cite{bianchi_coherent_2010,rovelli_covariant_2014}. Notice that we have used a slightly different convention of notation than the one usually taken in the literature. The phase factor of each link is included in the matrices $u_l$ and $v_l$. This simplifies the numerical calculations. Moreover, the notation $ \sum_{j_l} \prod_{l=1}^{N_L}$ must be understood as follows: for a given network, the state is given by a sum of terms labeled by the spin numbers associated with the links of the network (hence the notation $\sum_{j_l}$), and for a given set of spin numbers, its contribution to the sum is a product of terms associated with the links (hence the notation $\prod_l$).

Finally, we can define the scalar product between two gauge-invariant complexifier coherent states:
\begin{equation}
\braket{\psi^\zeta_{[g_l ']}}{\psi^\zeta_{[g_l]}} = \int_{SU(2)^{N_l}} \prod_{l=1}^{N_L} dU_l \overline{\psi^\zeta_{[ g_l ']} (U_1,...,U_{N_L}) } \psi^\zeta_{[ g_l ]} (U_1,...,U_{N_L}) .
\end{equation}
For additional details on gauge-invariant complexifier coherent states, we refer to~\cite{bahr_gauge-invariant_2009,stottmeister_coherent_I_2016,stottmeister_coherent_II_2016,stottmeister_coherent_III_2016}  Based on these definitions, we define the boundary term of the optimal control action:
\begin{equation}
S_B = \ii \log \left(
 \frac{\braket{\psi^\zeta_{[g_l ']}}{\psi^\zeta_{[g_l]}}}
 {\sqrt{
 \braket{\psi^\zeta_{[g_l ]}}{\psi^\zeta_{[g_l]}}
 \braket{\psi^\zeta_{[g_l ']}}{\psi^\zeta_{[g_l']}}
 }}\right).
 \label{eq:boundary_1}
\end{equation}

Similarly to the boundary term used with the harmonic oscillator, Eq.~\eqref{eq:boundary_1} is also related to the distance between two states, but in a more subtle way: the scalar product $\braket{K^\zeta_{g_l '}}{K^\zeta_{g_l}}$ is related to the $SL(2,\setC)$ geodesic lengths between $g_l$ and $g_l'$~\cite{bahr_gauge-invariant_2009}.

As a final remark, we note that it is also possible to consider the reduced gravitational state of a small portion of space. Reduced states play a key role in open quantum systems. They are also a relevant quantity to consider in experiments. A precursory study of such state is given in the appendix \ref{sec:density_matrix}.

\section{Optimal control path integral of general relativity}
\label{sec:path_integral_GR}

We can now proceed to the path integral quantization of general relativity using the optimal control formalism. For that purpose, we have to input the action $S_{tot} = S_B + S$ in a path integral. There is, however, a subtle point concerning the constraints $C_A$. They are non-dynamical and there is no associated state propagation. If these constraints are kept in the action, and if we integrate over $\lambda^A$, we obtain 7 additional Dirac distributions, and hence the path integral is a divergent generalized function (see e.g. \cite{colombeau_elementary_2011} for an introduction concerning generalized non-linear functions). This divergence is, however, non-physical and it is only the result that these constraints are considered on the same level as dynamical constraints. To avoid such a problem, we transform these constraints in a specific integration measure (see \cite{freidel_ponzano-regge_2004} for a similar discussion in the spinfoam setting):
\begin{equation}
\hat W = \int \Mc D E_i^a(x) \Mc D A^i_a(x) \Mc D \Pi_i^a(x) \Mc D P^i_a(x) ~~
 \Id_{\{C_A =0 \}}(E_i^a,A^i_a) ~~ e^{- \ii S'},
\label{eq:path_integral_GR_conti}
\end{equation}
where $S'$ is given by $S'=S_B + S$ , with $\lambda^A = 0$, and $\Id_{\{C_A=0\}}(E_i^a,A^i_a)$ is the indicator function such that $\Id_{\{C_A =0 \}}(E_i^a,A^i_a) = 1$ if all constraints $C_A = 0$ are verified, and  $\Id_{\{C_A=0 \}}(A_i^a,E^i_a) = 0$ otherwise. In practice, we should work with an approximation of the space-time manifold, using a finite number of points. To ensure well-defined discretized dynamics, one could use a weaker constraint, for example $|C_A|< \varepsilon$. These constraints can be imposed at each time step, or only on the boundary.

After the discretization on a lattice, the path integral \eqref{eq:path_integral_GR_conti} becomes.
\begin{equation}
\begin{split}
\hat W \propto \int \prod_{x,t} & d E_i^a(x,t)  d A^i_a(x,t) d \Pi_i^a(x,t)  d P^i_a(x,t)  \\
& \times \Id_{\{C_A =0 \}}(E_i^a(x,t),A^i_a(x,t)) ~~ e^{- \ii S'_{lattice}}.
\end{split}
\label{eq:path_integral_GR_discret}
\end{equation}
This integral is exactly in the same form as \eqref{eq:path_integral_delta}. Then, an integration over $\Pi_i^a(x,t)$ and $ d P^i_a(x,t)$ leaves us with:
\begin{equation}
\begin{split}
\hat W  = \int  \prod_{x,t} & d\mu ( E_i^a(x,t)  d A^i_a(x,t))  e^{- \ii S_B} \\
&\times \delta_0 (\Delta t [E_i^a (x,t) - \Mc G_i^a]) \\
&\times \delta_0 (\Delta t [ A_a^i (x,t) - \Mc F_a^i])
\end{split}
\label{eq:path_integral_GR_delta}
\end{equation}
with,
\begin{equation}
d\mu \left( E_i^a(x,t)  d A^i_a(x,t)\right) = 
 d E_i^a(x,t)  d A^i_a(x,t) ~ \Id_{\{C_A =0 \}}(E_i^a(x,t),A^i_a(x,t)) .
\end{equation}
Surprisingly, Eq.~\eqref{eq:path_integral_GR_delta} has a similar structure as spinfoams transition amplitudes \cite{baez_introduction_1999,perez_spin-foam_2013,rovelli_covariant_2014} $\int_{G^N} \prod_{e=1}^N dU_e \prod_f \delta (\prod_{e \in f} U_e)$, with $G$ the gauge group of the theory, $U_e \in G$ the holonomy associated with the edge $e$ of the spinfoam graph, and $f$ is a face, composed of several edges. However, the optimal control formalism offers a straightforward implementation of the cosmological constant. This is a non-trivial point in spinfoam theory.

From the material presented in Sec.~\ref{sec:Path integral quantization}, we easily deduce that the propagator $\hat W$ propagates semi-classical states, and returns the scalar product with these semi-classical states:
\begin{equation}
\bra{E'(t),A'(t)} ~\hat W(t)~ \ket{E(0),A(0)}\\
 = \exp^{- \ii S_B(E'(t),A'(t),\hat W^{-1}_t E(t), \hat W^{-1}_t A(t))}
 \label{eq:transition_amplitude}
\end{equation}
 The propagation of an arbitrary quantum state is possible with a mapping between coherent states and another basis of quantum states. 
 
Notice the strong similitude with the propagation of a standard quantum field in the Hamiltonian framework. The propagator propagates a classical solution in time, without any changes in the classical dynamics. For example, consider an electromagnetic wave, such has a plane wave. The Schrodinger equation evolves the plane wave (by multiplying the solution with a factor $e^{\ii \omega t}$) exactly as Maxwell's equations do. A similar thing happens for any solution of Maxwell's equations. Once a classical solution is known, it is possible to define the corresponding coherent state, and other non-classical states \cite{gardiner_quantum_2004}. Here, the same thing happens for a classical solution of Einstein's equations. 

The main difficulty of the approach is therefore to solve Einstein field equations. This is a very arduous task in 4D, even in numerical relativity \cite{baumgarte_numerical_2010}. Because we cannot expand the field variables in a plane wave basis, we cannot generate algorithmically the entire space of classical solutions. The only thing we can do (for the moment) is to find a set of solutions sufficiently large in order to describe some relevant physical effects with a good accuracy. This can give us a landscape of possible quantum mechanisms, and we can estimate some physical observables. The obvious drawback of this approach is that nothing guarantees that an important classical solution is missing in the calculation of the dynamics.

To fix the idea, we can work with the following inclusion of sets:
\begin{equation}
\Mc C' \supset \Mc C \supset \mathbf{C} \supset \mathsf{C},
\end{equation}
with $\mathbf{C}$ the ensemble of a solution of Einstein's equations, and $\mathsf{C}$ the ensemble of known solutions of $\mathbf{C}$. This set can be used to define a subspace of the quantum Hilbert space.

At this stage, one can ask where quantum mechanics is hidden? Is this formalism restricted to describe only semi-classical state? In fact no, and all non-classical effect are described by the boundary term that represents the scalar product. Non-classical effects can be induced by the coupling with matter fields.

\section{Application to black hole tunneling}
\label{sec:BH_tunneling}

By means of application, we consider the problem of transition of geometry from a black hole (B.H.) toward another black hole or toward a white hole (W.H.) This effect is usually called black-hole tunneling in the spinfoam formalism~\cite{christodoulou_planck_2016,dambrosio_end_2020,rovelli_planck_2014,barrau_planck_2014}. For this paper we consider a simple toy model for which the  effect is given by the transition amplitude between two geometries. Hence, we have to evaluate Eq.~\eqref{eq:transition_amplitude} between two B.H. (or W.H.) states.

For that purpose, we choose $\mathsf{C}$ to be the ensemble of Schwarzschild geometries labeled by the Schwarzchild radius $r_S$, the sign $s$ of the extrinsic curvature \cite{christodoulou_planck_2016}, and a time $\tau$. We present the calculations in Lemaitre coordinates, but other coordinate systems can be used as well. With these coordinates the line element is:
\begin{equation}
ds^2 = - d\tau^2 + \frac{r_S}{r} d\rho ^2 + r^2 d\theta ^2 + r^2 \sin ^2 \theta d\phi ^2,
\end{equation}
with,
\begin{equation}
r= \left( \frac{3}{2} (\rho - \tau)\right)^{2/3} r_S^{1/3}.
\end{equation}
The extrinsic curvature $K_{ab}$ for a $\tau = Cst$ hypersurface is given by a diagonal matrix:
\[
K_{ab} = \frac{2 s r^2}{3(\rho-\tau)}\text{diag}\left( \frac{2}{9(\rho-\tau)^2},-1,-\sin^2 \theta\right).
\]
We are then able to determine the "Gravitational electric field":
\[
E_i^a = \sin \theta .\text{diag}\left( r^2 , \sqrt{r r_S}, \sqrt{r r_s \text{csc}^2 \theta}\right),
\]
with $ \text{csc} \theta$ the cosecant function, and the Asketar connection:
\[
A_a^i = \left(\begin{array}{ccc}
\frac{ s \beta r_S}{2 r^2} & 0 & 0 \\ 
0 & -\frac{ s \beta r_S}{\sqrt{r r_S}} & -1 \\ 
-\frac{1}{2}\sin(2\theta)\text{csc} \theta  & \frac{1}{2}((\text{csc}\theta )^{-1} + \sin\theta) & -\frac{ s \beta\sqrt{r_S}}{\sqrt{r .\text{csc}^2\theta}}
\end{array}  \right)
\]

From these classical variables, we need to construct quantum states. Following the section~\ref{sec:GR}, we discretize the space-like submanifolds at $\tau = Cst$ by a set of $N$ points, for which we attach a couple $(A_a^i, E_i^a)$. Singular points must not be included in the set of points. As a result, the singularity is removed if one tries to reconstruct the discretized manifold from these data, but the singularity is recovered in the limit when an infinity of points are used to cover the entire hypersurface.

It has been argued for a long time that black hole singularities are removed in loop quantum gravity with a bounce. Recent studies \cite{bojowald_black-hole_2020} suggest that this may be a non-physical effect that breaks covariance. Since, it has been imagined that singularity is avoided with some process involved during the collapse (with, for example, a signature change). Hence, to solve the singularity problem completely, one should have a clear understanding of black holes formations. From this point of view, a Schwarzchild black holes is not only a "steady-state" geometry, but an approximated "steady-state" geometry. Therefore, the corresponding quantum state is only the approximation of a black hole quantum state, that can be used only for a first look at the quantum phenomenology. Also, the quantum interaction with matter fields may play a key role in the definition of a realistic black hole state. This issue is neglected in the following.
\begin{figure}[h]
\begin{center}
\includegraphics[width=\textwidth]{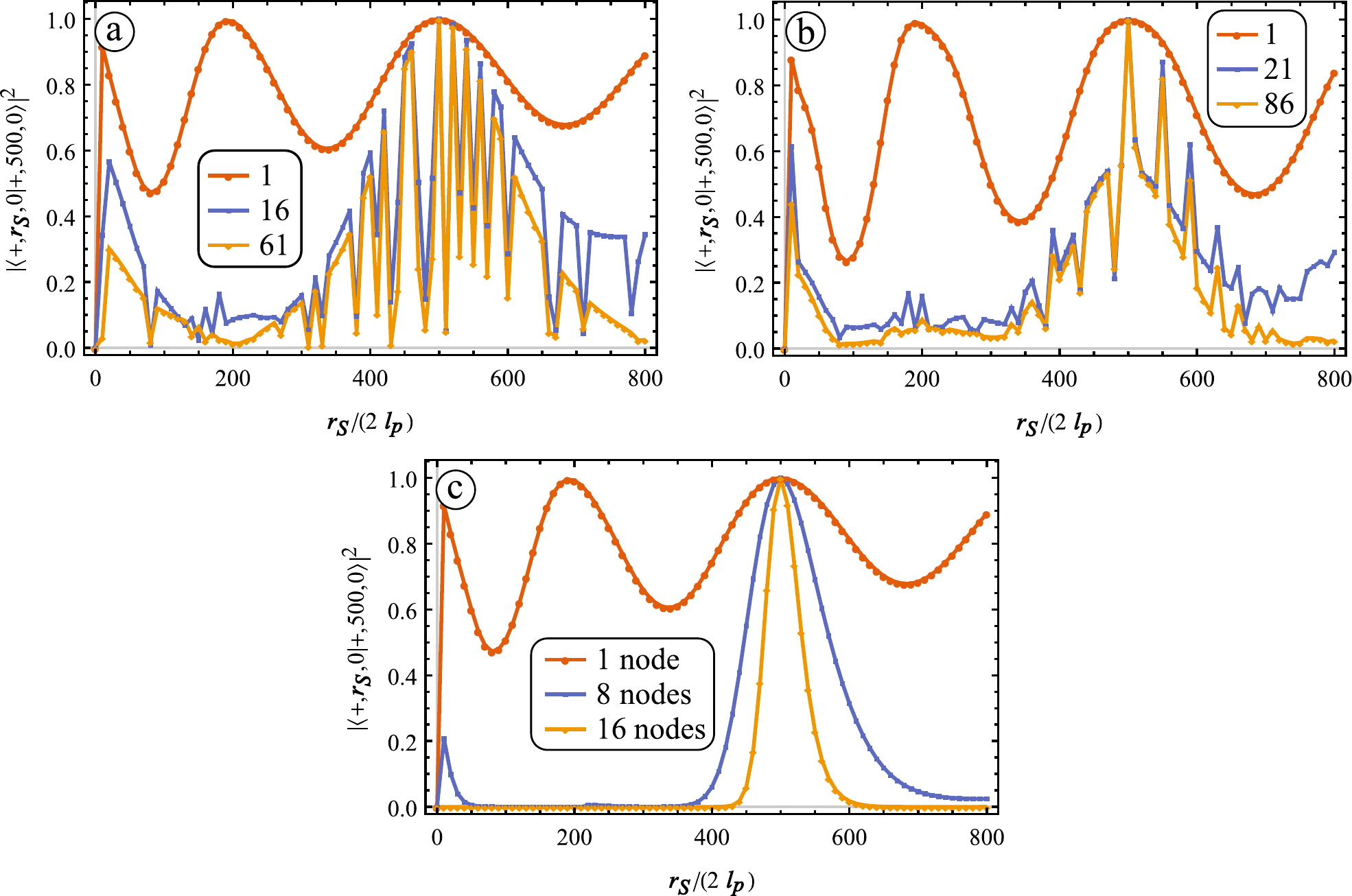}
\end{center}
\caption{B.H. to B.H. transition probability from $\ket{+,500,0}$ to $\ket{+,r_S,0}$ for different spin-network configurations. (a) Single node spin-network. (b) spin-network with two nodes connected by one link. (c) Ensemble of unconnected nodes. For the subfigures (a), and (b), each curve corresponds to a different number of graphs taken into account in the evaluation of the scalar product. The graphs are selected in order to keep only the ones with the largest contributions. As an example, for the subfigure (a), the orange curve is computed with $\{j_l\} = \{\{2,2,2,2,2,2\}\}$, and the blue curve is computed with $\{j_l\} = \{ \{2,2,2,2,2,2\}, \{\tfrac{3}{2}, \tfrac{3}{2},2,2,2,2\},... \}$. For the subfigure (c), only the graph with the highest contribution is kept, which means that the orange curve of (a) is the same as the orange curve of (c).}
\label{fig:BH_to_BH}
\end{figure}
\begin{figure}[h]
\includegraphics[width=\textwidth]{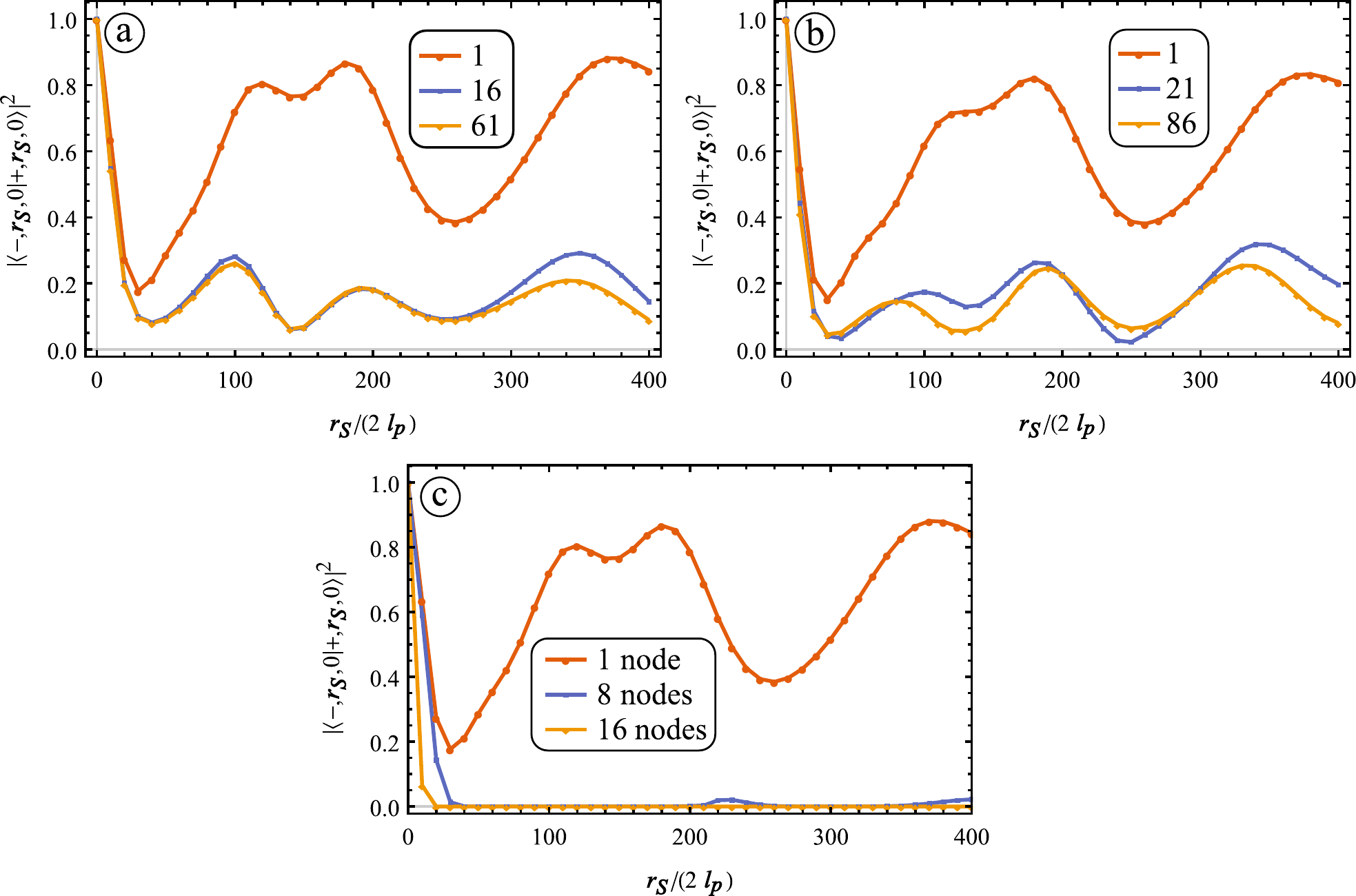}
\caption{Same as Fig.~\ref{fig:BH_to_BH}, but for the transition of geometry between a black hole state $\ket{+,r_s,0}$ to a white hole state $\ket{-,r_s,0}$.}
\label{fig:BH_to_WH}
\end{figure}

\begin{table}[t]
\begin{tabular}{|c|c|c|}
\hline
\includegraphics[width=1.5cm]{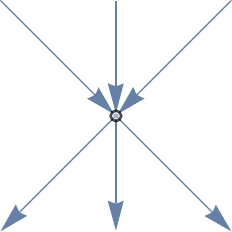} & \includegraphics[width=1.5cm]{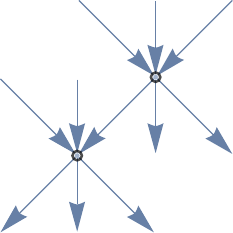} & \includegraphics[width=1.5cm]{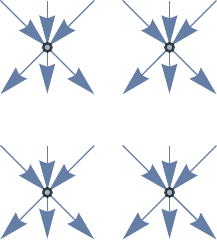} \\ 
a) A single node. & b) Two nodes with one common link. & c) Several unconnected nodes.\\
\hline
\end{tabular} 
\caption{Different types of networks considered to approximate the gravitational states.}
\label{tab:types_of_graphs}
\end{table}
To evaluate the probability transition, we have to deal with spin-networks with an infinite number of nodes, and we have to sum over an infinite number of spin labels. In practice, we have to restrict the analysis to a very small number of nodes, and to sum over the most relevant graphs. By increasing both the number of graphs and the number of nodes, we can extrapolate the result and deduce qualitative limit properties. Details concerning the numerical methods used to evaluate the scalar product between two coherent states are given in Appendix~\ref{sec:numerical_methods}. The networks considered in this paper are given in Tab.~\ref{tab:types_of_graphs}. 
In all cases, nodes have 6 links. Due to the network truncation, they may not be explicitly connected to other nodes. While the first two graphs are rather obvious to increase step-by-step the complexity of the spin-network, a comment on the last case seems necessary. Eq.~\eqref{eq:approx_coherent_state} tells us that the state is a product of terms associated with nodes and links. As an approximation, we can neglect all other entities surrounding a given region of the network, but we can also consider several unconnected regions. Then, we simply have to take the product of all the amplitudes associated with these regions. With this trick, we can explore a larger portion of space, where the gravitational field has very different strengths. Note that the coordinates $(\rho,\theta,\phi)$ of the nodes are kept fixed. While this is in general not necessary, we make this assumption in order to simplify the analysis of the transition probability when the size of the spin-network grows. 

In addition to these considerations, we have to choose a value of $\zeta$ (the parameter defining the coherent state). The semi-classical limit is obtained for $\zeta \rightarrow 0$. In this limit, the state is distributed over many different graphs with very large spins. This may render the numerical calculations extremely difficult. Then, it is necessary to choose the largest value of $\zeta$ as possible. With this constraint in mind, we have fixed $\zeta = 1/\sqrt{2}$, and we have adapted the size of the discretization grid in order to obtain $r_l \approx 1.7$. This gives us a maximum of $e^{-\zeta j_l (j_l +1)/2} d_{j_l} e^{r_l j_l}$ around $j_l = 2$. All the data required to reconstruct the boundary state are given in the supplementary materials.

Figure \ref{fig:BH_to_BH} and \ref{fig:BH_to_WH} show examples of transition probability, for respectively a B.H. to B.H. transition, and a B.H. to W.H. transition. These transitions are given by $|\braket{+,r_S,0}{+,500,0}|^2$ and $|\braket{-,r_S,0}{+,r_S,0}|^2$.  In the first case, the initial black hole has a radius $r_S/(2 l_P) =500$ (with $l_P$ the Planck length), and the radius of the final black hole is allowed to change. In the second case, the radius is identical for both the initial and final states but the sign of the extrinsic curvature $s$ is changed. In these examples, the time is fixed to $\tau = 0$.

In both cases we observe that increasing the number of nodes or increasing the number of graphs taken into account in the sum leads to a transition probability extremely peaked around the initial state. The fast oscillation observed in subfigures~\ref{fig:BH_to_BH} (a) and (b) are induced by a kind of interference between the graphs. While it is not shown explicitly in this paper, similar results have been observed for the transition between several time slices. Then, by extrapolating these results, we arrive at the conclusion:
\begin{equation}
|\braket{s',r_S',\tau '}{s,r_S,\tau}|^2 =\left\{ \begin{array}{cc}
\delta_{ss'} \Id_{\{r_S\}} (r_S') \Id_{\{ \tau \}}(\tau') &, ~r_s \neq 0 \\ 
\Id_{\{0\}}(r_s') & ,~ r_s = 0
\end{array} \right.
\label{eq:transition_proba_BH_to_BH}
\end{equation}
When $r_S = 0$, black hole ($s=+$) and the white hole ($s=-$) solutions are identical. In this latter case, the system is also time invariant.

With Eq.~\eqref{eq:transition_proba_BH_to_BH}, the tunneling effect is impossible or at least extremely improbable. However, such a phenomenon could be enhanced by the interaction with matter. During the black hole formation, we might have a strong entanglement with the matter field that produces a squeezed gravitational state. This state might have a large spreading and it can make possible a kind of tunneling. This issue can be addressed only with an accurate modeling of the B.H. formation that takes into account the quantum interaction between gravity and matter fields. This issue is investigated in \cite{achour_towards_2020} with the framework of Loop Quantum Cosmology, and several key ingredients could be adapted to the approach of this paper.

\section{Conclusion}
\label{sec:conclusion}

The aim of this paper is to present a novel path integral formalism for general relativity, based on the extended framework of optimal control theory. Using an extended configuration space of classical variables, we are able to construct a new Lagrangian for gravity. Variations of the action functional with respect to the adjoint variables gives (in standard optimal control theory) a set of admissible optimal trajectories. Here, this corresponds to the set of space-time satisfying Einstein field equations. 

The time discretization of the dynamical system allows us to define rigorously the propagator. The physical scalar product of the Hilbert space is implemented in the theory using a boundary term that describes a distance measure in the space of classical solutions. This boundary term is analogous to the one introduced in optimal control theory to relax the target state constraint. The formalism is flexible enough to define the propagator with any set of classical variable expressed in a 3+1 formalism. Explicit calculations are made using Ashtekar variables, in order to define the boundary term rigorously with the LQG scalar product.

By means of an example, the tunneling problem of a black hole is tackled using a subset of the Hilbert space generated by Schwarzchild geometries. This allows us to determine numerically the transition probability from a black hole toward another black hole, or toward a white hole. Numerical calculations suggest that such transitions are impossible (or at least, extremely improbable), but additional mechanism coming from the interaction with matter fields may change the result.

The theory presented in this paper is only at an early stage, and many points are left unexplored. Future studies shall focus on the relation between the Pontryagin Hamiltonian for gravity and the ADM Hamiltonian. It must also be clarified the role that Pontryagin Hamiltonian can play in the canonical quantization. Also, a clearer link with Oeck's general boundary formalism~\cite{oeckl_general_2005} could be interesting. Finally, we point out that the coupling with matter fields is straightforward in this formalism. We have not discussed this topic since it must be addressed simultaneously with decoherence and quantum measurements.

\section*{Acknowledgements}
The author acknowledges David Viennot for useful discussions and helpful comments.

\appendix
\section{Hamiltonian operator}
\label{sec:hamiltonian operator}

This appendix gather some preliminary results concerning the Hamiltonian quantization using the Optimal Control formalism.
Main ideas are first introduced using the harmonic oscillator, and the case of gravity is considered in a second step.

Given a function $F \in \Mc C'$, its time derivative is given by the Poisson-bracket:

%
\begin{equation}
d_t F= \{ F ,   H_{OC} \},
\end{equation}
with,
\begin{equation}
\{F, H_{OC} \} = \sum_i \frac{\partial F}{\partial x^i} \frac{\partial   H_{OC}}{\partial p_{x^i}} - \frac{\partial F}{\partial p_{x^i}} \frac{\partial   H_{OC}}{\partial x^i}.
\end{equation}
The quantum state is assumed to be a function on $\Mc C$. Then, the Poisson-bracket is reduced to $\{\psi, H_{OC} \} = \sum_i \frac{\partial \psi}{\partial x^i} \frac{\partial   H_{OC}}{\partial p_{x^i}}$. Moreover, the Hamiltonian in the optimal control framework is of the form: $H_{OC} = \sum_i p_{x^i} f^i$, where $f^i = \dot x^i$. Then, we notice that the quantization rule $p_{x^i} \rightarrow  \ii \partial_{x^i}$ allows us to define an Hamiltonian operator $\hat H_{OC} = \ii f^i \partial_{x^i}$, such that:
\begin{equation}
 d_t \psi = - \ii \hat H_{OC} \psi.
\end{equation}
Precisely, for the harmonic oscillator, we have:
\begin{equation}
\hat H_{OC} = \ii \left( \frac{p}{m} \partial x - k x \partial_p \right)
\end{equation}
This operator is equivalent to the standard Hamiltonian operator in the holomorphic representation \cite{gardiner_quantum_2004}:
\begin{equation}
\hat H = \omega \alpha \partial _\alpha
\end{equation}
with $\alpha$ defined by equation \eqref{eq:def_alpha}. An explicit calculation gives us: $d_t \alpha = -\ii \hat H_{OC} \alpha = -\ii \hat H \alpha$, and more generally, for any holomorphic function $\psi$: $\hat H_{OC}\ket{\psi(\alpha)} = \hat H \ket{\psi(\alpha)}$. The optimal control Hamiltonian allowed us to construct straightforwardly the Hamiltonian operator in the holomorphic representation of the harmonic oscillator. We shall proceed similarly with gravity. From the results of Sec.~\ref{sec:GR} the Hamiltonian is:
\begin{equation}
\begin{split}
H_{OC} (x)= & P_a^i(x) \Mc G_i^a (x) + \Pi^a_i(x) \Mc F_a^i (x)  \\
& - \lambda^A (x) C_A (x),
\end{split}
\end{equation}
and the corresponding Hamiltonian operator is given by:
\begin{equation}
\begin{split}
\hat H_{OC} (x) = & \ii\left( \Mc G_i^a (x) \frac{\partial}{\partial E_i^a (x)} +  \Mc F_a^i (x) \frac{\partial}{\partial A^i_a (x)} \right. \\
& \left. - C_A(x) \frac{\partial}{\partial C_A (x)}\right) .
\end{split}
\end{equation}
Assuming that $\ket{ \psi}$ does not depend explicitly on $C_A$, the part $C_A \partial_{C_A}$ vanishes in $\hat H_{OC}$. Using operations on the loop space, it is possible to transform $\hat H_{OC}$ into an operator on loop quantum states (like in Refs.~ \cite{gambini_loops_1996,rovelli_quantum_2007}). This step requires some cautions, for the same reasons as the ones that make difficult the canonical quantization. However, using the link between coherent spin-networks and classical geometries could provide a welcome simplification. A detailed study of this Hamiltonian operator is left for another paper.

\section{Numerical methods}
\label{sec:numerical_methods}

Spin networks are objects with many degrees of freedom and the evaluation of the scalar product between different coherent spin-network states can be challenging. In this appendix, we provide several details concerning the numerical methods employed for the calculations. 

\textit{In the following, all computation times are given for a single core clocked at 3.3 Ghz. Numerical calculations are made using Mathematica.}

The numerical algorithm used for the calculation  is based on several key ingredients. First of all, the scalar product is evaluated using an approximation derived in the large $j$-limit. Using similar computation steps as the ones between Eq.~\eqref{eq:derivation_heat_kernel_approximation} and Eq.~\eqref{eq:approx_coherent_state}, we can derive an approximated expression for the scalar product between two heat kernels:
\begin{equation}
\begin{split}
\braket{K^\zeta _{h_a g_l h_b^{-1}}}{K ^\zeta_{h_a' g_l ' h_b'^{-1}}} & = \sum_{j_l} e^{-\zeta j_l(j_l+1)}d_{j_l} Tr^{j_l}\left( h_b g_{l}^\dagger h_a ^\dagger h_a' g_{l}'h_b'^\dagger\right) \\
& \approx  \sum_{j_l} e^{-\zeta j_l(j_l+1)}d_{j_l} e^{(r_l+r_l')j_l} 
 D^{j_l}_{j_l j_l} \left(  v_{l}'^\dagger h_b'^\dagger h_b v_{l}\right) D^{j_l}_{j_l j_l}\left(u_{l}^\dagger h_a ^\dagger h_a' u_{l}'\right).
\end{split}
\label{eq:braket_approx_1}
\end{equation}
where $h_a,h_b,h_a',h_b'$ are $SU(2)$ matrices associated with the nodes which are used to produce gauge invariant states, and $g_l,g_l'$ are rewritten using a Cartan decomposition $g_l = u_l.e^{r_l \sigma_z/2}.v_l^\dagger$, with $r_l\in \setR_+$ and $u_l,v_l \in SU(2)$.

In the last line, we observe that $u$ and $v$ are separated into different Wigner matrices. We are left with a product of terms, each one being associated with different nodes. Now, if we consider the full gauge coherent state, we have:
\begin{equation}
\braket{\psi_{[g_{l}]}^\zeta}{\psi_{[g_{l}']}^\zeta}  =\sum_{j_l} \int_{SU(2)^{2N_N}} \prod_{a=1}^{N_N} dh_a dh_a' 
 \prod_{l=1}^{N_L}  e^{-\zeta j_l(j_l+1)} e^{(r_l+r_l')j_l}d_{j_l} D^{j_l}_{j_l j_l}\left(w_{l}^\dagger h_a ^\dagger h_a' w_{l}'\right)
\label{eq:braket_approx_full}
\end{equation}
where $w_l$ corresponds to $u_{l}$ or $v_{l}$, depending on the orientation of the link. Then, the scalar product is a sum of terms labeled by all possible combinations of $j_l$. For each term, we have a link contribution $ e^{-\zeta j_l(j_l+1)} e^{(r_l+r_l')j_l}d_{j_l}$, and a node contribution given by the integral of a product of functions $ D^{j_l}_{j_l j_l}(...)$. This integral can be computed numerically using the following scheme:
\begin{enumerate}
\item For each matrix $g_l$ defining $\psi_{[g_{l}]}^\zeta$ and $\psi_{[g_{l}']}^\zeta$, determine $u_l,v_l,r_l$ such that $g_l =  u_l.e^{r_l \sigma_z/2}.v_l^\dagger$. This can be achieved easily using the build-in diagonalization function of Mathematica: \textit{Eigensystem}. When a matrix of $SL(2,C)$ is diagonalized numerically, the algorithm returns the matrix of eigen-vectors $P$ in the form of a matrix of $SU(2)$, and the eigenvalues are returned in the form $(e^{z  /2},e^{-z/2}),~ z \in \setC$, but if the matrix to diagonalize is a pure boost, we have $z \in \setR$. Using the fact that $g_l = H_l . U_l$, with $H_l$ defined in Eq.~ \eqref{eq:def_matrix_gl} and $U_l$ defined in Eq.~\eqref{eq:holonomy}, we can decompose $g_l$ as follows: $u=P^\dagger$, $r= z$, $v^\dagger = P.U_l$, where $P$ and $z$ are deduced from the diagonalization of $H_l$.

\item The second step is to determine the set of values ${j_l}$ such that $ \prod_l e^{-\zeta j_l(j_l+1)} e^{(r_l+r_l')j_l}d_{j_l}$ is sufficiently high. This allows us to reduce considerably the number of terms in the final evaluation of the spin networks. In practice, many terms have a weight of zero or a negligible weight. The choice of the most relevant graphs taken into account in the computation is discussed later in this appendix.

\item Compute the node amplitudes
\[
\int_{SU(2)^2} dh_a dh_a' \prod_l  D^{j_l}_{j_l j_l}\left(w_{l}^\dagger h_a ^\dagger h_a' w_{l}'\right).
\]
Since $SU(2)$ is a compact Lie group, $\int f(h h') dh = \int f(h^\dagger)dh = \int f(h)dh$. Then, we can drop an integral, and we are left with:
\begin{equation}
\int_{SU(2)} dh_a \prod_l  D^{j_l}_{j_l j_l}\left(w_{l}^\dagger h_a w_{l}'\right) 
=\int_{0}^{4 \pi}d \psi \int_{0}^\pi d\theta \int_0^{2\pi} d\phi \sin(\theta)  \prod_l \left[\left(w_{l}^\dagger h_a(\psi,\theta, \phi) w_{l}'\right)_{1,1} \right]^{2j_l}.
\label{eq:node_amplitude}
\end{equation}
In the second line we have introduced the matrix element $(1,1)$ of $w_{l}^\dagger h_a w_{l}'$, and the  Euler angles  $(\psi,\theta,\phi)$ of $h_a$. Two options are available. The first option provides an exact result (up to the numerical precision). The idea is to expand $\sin \theta \prod_l\left[\left(w_{l}^\dagger h_a w_{l}'\right)_{1,1} \right]^{2j_l}$ into a polynomial of the form:
\begin{equation*}
\sum_{k,l,..}c_{klmnop} \cos^k{\frac{\psi}{2}} \sin^l{\frac{\psi}{2}} \cos^m{\frac{\theta}{2}} \sin^n{\frac{\theta}{2}}  \cos^o{\frac{\phi}{2}} \sin^p{\frac{\phi}{2}}
\end{equation*}
with $c_{klmnop} \in \setC $. This step can be performed with repeated uses of the function \textit{CoefficientList}. Then, an exact evaluation is possible with the identities:
\[
2\int_0^{2\pi} \cos^n\psi \sin^m\psi d\psi = (1+(-1)^n)(1+(-1)^{n+m}) \frac{\Gamma((m+1)/2) \Gamma((n+1)/2)}{\Gamma((m+n+2)/2)} 
\]

\[
\begin{split}
2\int_0^{\pi/2} \cos^n\theta \sin^m\theta d\theta =  \frac{\Gamma((m+1)/2) \Gamma((n+1)/2)}{\Gamma((m+n+2)/2)} 
\end{split}
\]

\[
2\int_0^{\pi} \cos^n\phi \sin^m\phi d\phi = (1+(-1)^n) \frac{\Gamma((m+1)/2) \Gamma((n+1)/2)}{\Gamma((m+n+2)/2)} .
\]
\begin{figure}[t]
\begin{center}
\includegraphics[width=\textwidth]{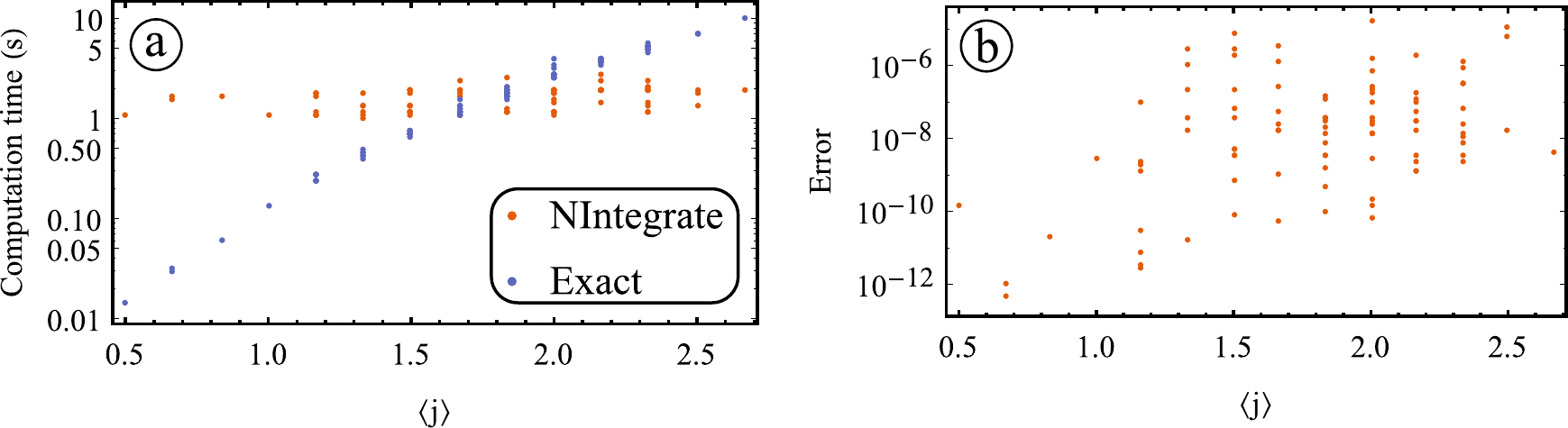}
\end{center}
\caption{a) computation time of Eq.~\eqref{eq:node_amplitude} as a function of $\langle j \rangle = \sum_{l =1} ^6 j_l/6$, for the two methods (we recall that the nodes considered in this study have 6-links). b) Error of the \textit{NIntegrate} method as a function of $\langle j \rangle$. Each point of these plots corresponds to an integration, for which all $j_l$, $w_l$, $w_l'$ are generated randomly.}
\label{figD1}
\end{figure}

The other method is based on a "brute force" numerical integration using the function \textit{NIntegrate} with the method \textit{MultiPeriodic}. This method is specifically designed for integration of highly oscillating functions. The two methods are compared in Fig.~\ref{figD1}. The exact method is particularly efficient for small values of $j_l$, but the computation time increases exponentially. On another side, the numerical estimation with \textit{NIntegrate} has a stable computation time and a small error $\lesssim 10^{-6}$(the error is defined by the absolute value between the approximated result and the exact one).  We observe that the exact method is faster for $\langle j \rangle = \sum_{l =1} ^6 j_l/6 <2$. Then, we can adapt the integration scheme as a function of $\langle j \rangle$.
\end{enumerate}

As outlined above in this section, all labeled graphs do not contribute equivalently in a coherent state. It can be sufficient to compute the scalar product using only a small number of graphs. This can reduce the accuracy of the computation, but this gives us qualitative behaviors. We describe below a list of properties that allows us to select the relevant graphs.

\begin{itemize}
\item The integration over $SU(2)$ at each node induces a selection rule for the different spin numbers. For the 6-valent intertwiner, we have the following condition: $\sum_l j_l \in \setN$.
\item For an arbitrary configuration, we can decompose each $j_l$ into an integer part and a half integer part: $j_l = n_l + \delta_l$ with $n_l \in \setN$, and $\delta_l = 0$ or $1/2$. If all spins in the graph have integer values, the sum of the spins at each node is an integer, and thus it is a valid graph. Non-trivial cases are given when there are $\delta_l \neq 0$. In fact, the validity of the graph depends only on the set of $\delta_l$, but not on the values of $n_l$. Hence, we can first determine the set of valid graphs with $j_l = \delta_l$, and then we can generate all other graph by adding to these solutions an integer $n_l$ to each $j_l$.
\end{itemize}

\section{Toward reduced gravitational states}
\label{sec:density_matrix}

\begin{SCfigure}
\includegraphics[width=7cm]{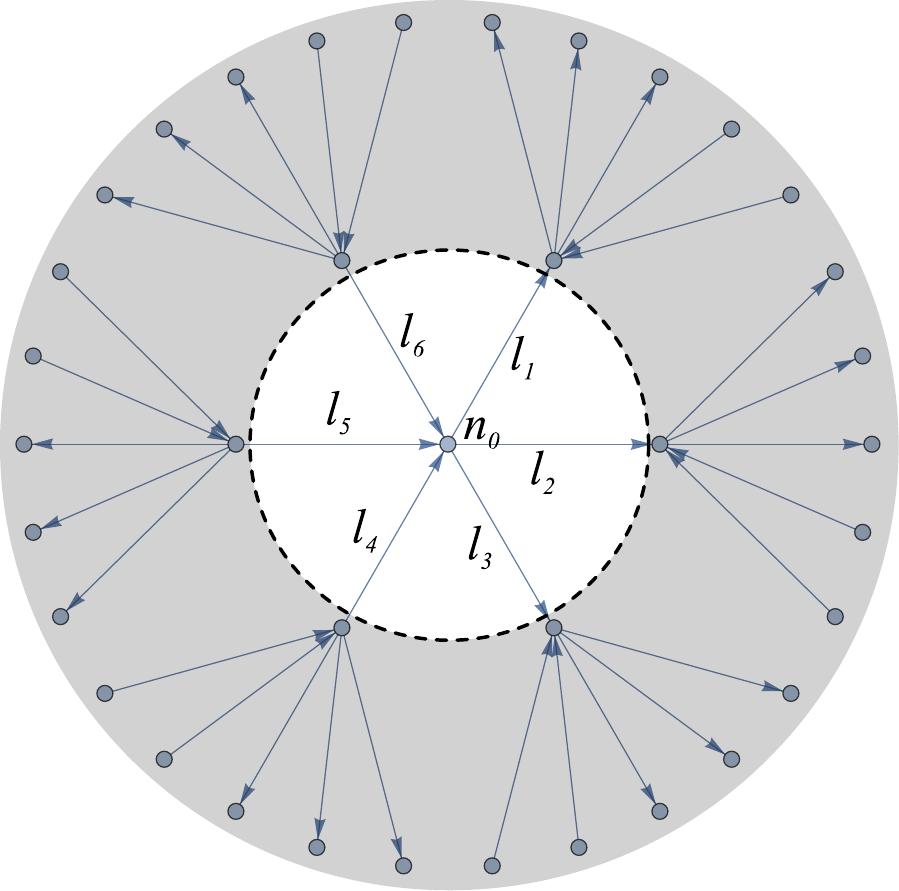}
\caption{Illustrative example of sub-graph induced by the partial trace over the links of a spin-network (for simplicity, only the links and nodes surrounding the interesting area are shown). The part of the graph in the gray area corresponds to the entities eliminated by the partial trace. The result is then a sub-graph with the node $n_0$, and links $l_1, ...,l_6$.}
\label{fig:sub_spin_network}
\end{SCfigure}
The quantum state $\ket{\psi}$ is an idealized concept that is, in practice, experimentally not accessible. Rather than $\ket{\psi}$, we may have access to the reduced density matrix of the system (and the reduced density matrix may not be a pure state of the form $\ket{\psi}\bra{\psi}$). The reduced state consists of the density matrix describing the state of "everything" where we have made a partial trace over all degrees of freedoms that do not describe the system of interest. Introducing reduced states in quantum gravity is interesting to model more realistic systems.

The density matrix for spin network states can be defined as follows. Given any set of spin networks, we can define $\ket{\psi (U_1,..,U_{N_L})} = \sum_i \psi_i \ket{i (U_1,..,U_{N_L})}$ where $\ket{i}$ is a spin-network, function of $N_L$ variables $U_l\in SU(2)$. Then, a density matrix is simply given by:
\begin{equation}
\rho (U_1,..,U_{N_L},U_1',...,U_{N_L}') = \sum_{i,i'} c_{i,i'}\ket{i (U_1,..,U_{N_L})}\bra{i'(U_1',..,U_{N_L}')},
\end{equation}
such that $\sum_i c_{ii} = 1$. Since nodes and links describe geometric quantities, we can define the partial trace over space degrees of freedoms by integrating over $SU(2)$ for each link outside the region of interest:
\begin{equation}
\rho_S =  \int_{SU(2)^{N_L - N_L'}} \prod_{l=1}^{N_L - N_L'} dU_l~\delta(U_l.U_l'^{-1}) ~ \rho (U_1,..,U_{N_L},U_1',...,U_{N_L}').
\label{eq:def_partial_trace}
\end{equation}
Remark that the integration over $U_l '$ is in fact hidden in the definition of $\bra{i}$.
The result of this partial trace leads to a kind of spin-network with links connected to a single node. Hence, we are forced to work with a new kind of state defined on open graphs. These states are called \textit{sub-spin-networks}. The idea is illustrated in Figure~\ref{fig:sub_spin_network}.

Eq.~\eqref{eq:approx_coherent_state} allows us to derive an approximated reduced density matrix of a gravitational state $\rho = \ket{\psi^\zeta_{[g_l]}} \bra{\psi^\zeta_{[g_l]}}$. Using the partial trace defined in Eq.~\eqref{eq:def_partial_trace}, we deduce:
\begin{equation}
\rho_S = \sum_{j_l, j_l'} c(\{j_l\},\{j_l'\}) \prod_{l,l'} e^{-\zeta j_l (j_l +1)/2} d_{j_l} e^{-\zeta j_l' (j_l' +1)/2} d_{j_l'}  e^{r_l j_l}  e^{r_l' j_l'} \ket{\Psi_{j_l, u_l, v_l}} \bra{\Psi_{j_l', u_l', v_l'}},
\end{equation}
with $ c(\{j_l\},\{j_l'\})$, a term that depends on all the $j_l$ of a given graph (here, $l$ runs over all the links of the sub-network). This can be viewed as a coefficient measuring the coherence between two graphs. Its value depends on the links and nodes amplitudes of the region of the graph which is traced out. 

Under several assumptions we can estimate these coefficients for a given reduced state. We explore briefly the case illustrated in Fig.~\ref{fig:sub_spin_network}, for a reduced density matrix of a single 6-valent node. We assume that the node is a part of a bigger graph with 6 other nodes and 30 other links. This is still a sub-graph of a hypothetical larger graph, but we assume that the coefficients depend mostly on the nearest neighbors. They are estimated by first choosing a finite set of sub-graphs with link labels $\{j_1,...,j_6\}$. For each couple of sub-graphs, we generate randomly the spin-numbers of the rest of the graph (i.e. the values of the 30 other $j_l$), and we compute the amplitude $\prod_{k=1}^{30} e^{-\zeta j_k (j_k +1)} d_{j_k} e^{2 r_k j_k}  \braket{\Psi_{j_k, u_k, v_k}}{\Psi_{j_k, u_k, v_k}}$. Without an explicit knowledge of the state in the traced region, we fix $r_k = r$, with $r$ a chosen value, and we assume $ \braket{\Psi_{j_k, u_k, v_k}}{\Psi_{j_k, u_k, v_k}} = 1$ or $0$ if the configuration is non-physical (the sum of the $j_l$ at each node must be an integer). This may be improved by introducing a more realistic estimation of $\braket{\Psi_{j_k, u_k, v_k}}{\Psi_{j_k, u_k, v_k}} $, for example, by using a random number with a specific probability distribution. However, our first investigations suggest that this does not change the results significantly. The coefficients are deduced by summing the amplitude of many (ideally an infinite number) of random graphs. The results can be normalized to obtain $\text{Tr}(\rho_S) = 1$.

\begin{figure}[t!]
\begin{center}
\includegraphics[width = 6cm]{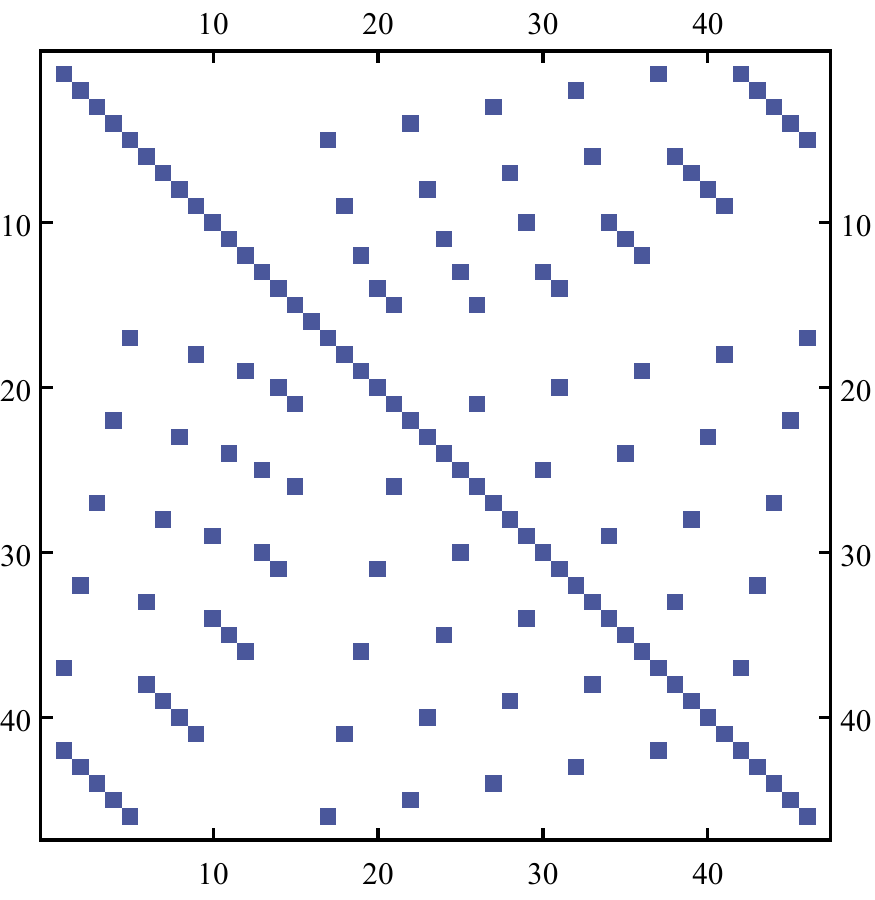}
\end{center}
\caption{Numerical estimation of $c(\{j_l\},\{j_l'\})$ for a 1-node reduced density matrix. We show only the sign of the coefficient (0 or 1, given respectively by white or blue pixels). While it is easy to deduce if a coefficient is different from 0, it is, however, very difficult to estimate its exact value (the convergence of the Monte-Carlo methods is very slow). Based on our numerical observation we conjecture that, for the case presented here, they are almost all identical. Calculations are made using $r = 1.5$ and a set of $46$ graphs with spin numbers $j_l$ in the range $[\tfrac{3}{2}, \tfrac{5}{2}]$.}
\label{fig:correlation_coeff}
\end{figure}

An example of coefficients is given in Fig.~\ref{fig:correlation_coeff}. Due to the non-zero diagonal elements and the sparse structure of non-diagonal elements, we see that the reduced density matrix is in a mixed state. Using the fact that the sum of the $j_l$ at each node must be an integer, we can show that the statistical mixture is induced by half integer spins while some of the coherence is kept by integer spins. The reduced density matrix formalism may be interesting to compute mean values of observable, but it is not well adapted to the problem studied in section~\ref{sec:BH_tunneling}. Probability transition of mixed states must be computed using the quantum fidelity, and this quantity is difficult to compute with coherent states.

\newpage
\bibliography{biblio}

\end{document}